\documentclass[]{aa}  

\usepackage{graphicx}
\usepackage{epstopdf}
\usepackage{txfonts}
\usepackage{ulem} 
\usepackage{amsmath,amsfonts,amssymb}
\usepackage{color}

\begin{document}

   \title{The chemical composition of planet building blocks as predicted by stellar population synthesis}

   \titlerunning{The chemical composition of planet building blocks as predicted by stellar population synthesis}
   \author{N. Cabral \inst{1}, N. Lagarde \inst{1}, C. Reyl\'e \inst{1}, A. Guilbert-Lepoutre \inst{1} and A.C. Robin \inst{1}}
   \institute{Institut UTINAM, CNRS UMR6213, Univ. Bourgogne Franche-Comt\'e, OSU THETA Franche-Comt\'e-Bourgogne, Observatoire de Besan\c con, BP 1615, 25010 Besan\c con Cedex, France \\
              \email{nahuel.cabral@utinam.cnrs.fr}
                }

   \date{Received  xx xx xx / Accepted xx xx xx}
 
   \titlerunning{The chemical composition of planet building blocks as predicted by stellar population synthesis}  
   \authorrunning{Cabral et al.}

  \abstract 
   {Future space missions (TESS, CHEOPS, PLATO and JWST) will improve considerably our understanding of the formation and history of planetary systems providing accurate constraints in planetary radius, mass and atmospheric composition. Currently, observations show that the presence of planetary companions is closely linked to the metallicity and the chemical abundances of the host stars.}
   {We aim to build an integrated tool to predict the planet building blocks composition as a function of the stellar populations, for the interpretation of the ongoing and future large surveys. The different stellar populations we observe in our Galaxy are characterized by different metallicities and alpha-element abundances. This paper investigates the trends of the expected planet building blocks (PBB) composition with the chemical abundance of the host star in different parts of the Galaxy.}
   {We synthesize stellar populations with the Besan\c{c}on Galaxy model (BGM) which includes stellar evolutionary tracks computed with the stellar evolution code STAREVOL. We integrate to the BGM a simple stoichiometric model already published by Santos and coll. (2017) to determine the expected composition of the planet building blocks.}
   {We determine the expected PBB composition around FGK stars, for the four galactic populations (thin and thick disks, halo and bulge) within  the Milky Way. Our solar neighborhood simulations are in good agreement with the recent results obtained with the HARPS survey for $f_\text{iron}$, $f_\text{w}$ and the heavy mass fraction $f_\text{Z}$. We present evidence of the clear dependence of $f_\text{iron}$ and $f_\text{w}$ with the initial alpha abundances [$\alpha$/Fe] of the host star. We find that the different initial [$\alpha$/Fe] distributions in the different galactic populations lead to a bimodal distribution of PBB  composition and to an iron/water valley separating PBB with high and low iron/water mass fractions.}
   {We linked host star abundances and expected PBB composition in an integrated model of the Galaxy. Derived trends are an important step for statistical analyses of expected planet properties. In particular, internal structure models may use these results to derive statistical trends of rocky planets properties, constrain habitability and prepare interpretation of on-going and future large scale surveys of exoplanet search.}
   

   \maketitle

%
\section{Introduction}

Exploring correlations between planet and star properties provide key constrains on planet formation models. Space-based (\textit{Kepler}, CoRoT) and ground-based exoplanets surveys (HARPS) have shown many dependences between the exoplanets properties and the physical properties of the host star. Moreover, TESS, CHEOPS, and JWST are expected to bring huge improvements in our characterizing of planets. The asteroseismic survey PLATO in particular will provide stellar properties with high accuracy \citep{Rauer2014,Miglio2017}. 

The main observed links between the host star chemical properties and its planet are:

\textbf{(i) The metallicity correlation with planet frequency.} Giant planets are observed more frequently in high-metallicity stars \citep[see e.g.][]{Santos2004, FischerValenti2005}, underlying the crucial role of metallicity in planetary formation processes. This correlation is not observed for small planets \citep{Sousa2011, Buchhave2012}. \cite{Zhu2016} suggests that the null detection of the metallicity correlation with small planets may be due to observational bias (the combination of high small planet occurrence rate and low detection efficiency). Currently, the correlation with giants and the very weak correlation with small planets, if exists, appears to be well explained by the core accretion model. In this model, giant formation efficiency is dependent on the metallicity, while small planets form even around very metal-poor stars \citep[see e.g.][]{IdaLin2004}. We note that the Tidal Downsizing Model is also reproducing both correlations \citep[][]{NayakshinFletcher2015}. Moreover, we should mention that there is a possible positive correlation between the masses of heavy elements in the planets and the metallicities of their parent star that is still to be confirmed \citep{Guillot2006,Thorngren2016}.

\textbf{(ii) The metallicity correlation with planet orbital distribution.} Different observational and theoretical studies illustrate the links between the metallicity and the orbital distributions. \cite{BeaugeNesvorny2013} found that small planets ($R_\text{p}$<4 $R_\oplus$) orbiting around metal-poor stars present larger periods (P>5 days) than small planets orbiting around metal-rich stars. The study of \citet{Adibekyan2013} suggests that the lack of small planets around metal-poor stars at short periods extend up to 4 Jupiter masses. Recently, by analyzing 212 close-in planets from the APOGEE \citep{Majewski2017} program, \cite{Wilson2018} showed that small planets with an orbital period below 8.3 days have a more metal-enriched host star. Moreover, \citet{DawsonMurray-Clay2013} showed that close-in giant planets orbiting [Fe/H]<0 host stars generally have lower eccentricities than those orbiting metal-rich stars. In this line, \cite{Buchhave2018}, found that cool Jupiters (a>1 AU) with low eccentricities (e<0.25) orbits on average metal-poor stars.

\textbf{(iii) The alpha abundance correlation with planet frequency.}
Observations suggest that the correlations between chemical stellar properties and exoplanets are not limited to the metallicity. The observed abundances of specific element ratios also show interesting correlations. Iron-poor stars hosting planets are found to preferentially present enhanced alpha elements composition \citep[see e.g.][]{Haywood2008,Haywood2009,Adibekyan2012a,Adibekyan2012b}. Interestingly, stars hosting low-mass planets (<30 M$_\oplus$) have higher Mg/Si abundance ratios than stars with giant planets (>30 M$_\oplus$) \citep[][]{Adibekyan2015}. This indicates the important role of Mg/Si ratio in the formation of these small planets. It also suggests that the heavy elements such as Mg could compensate for the lack of iron to form low-mass planets in metal poor host stars and could explain the absence of correlation of low-mass planets with [Fe/H]. As discussed by \citet{Adibekyan2015}, frequency of low-mass planets should correlate with refractory elements (as Mg, Si and Fe) and not necessarily with iron only.\\

As planets properties are observed to correlate with metallicity and specific elemental ratios of the host stars, the different stellar galactic populations could produce planets with very different properties. Indeed the stellar populations in the Milky Way  show different metallicities and $\alpha$-abundances due to their different way and epoch of formation and chemical evolution. The halo contains the more metal poor stars. The disk exhibits two sequences (thick and thin discs) where the thick disk stars are in general more metal-poor and alpha-enriched compared to thin disk stars \citep[see e.g.][]{Haywood2013}. The bulge presents a range in metallicities similar to the thin disk but a much larger spread in alpha abundances.

\citet[hereafter S17]{Santos2017} determined the expected chemical composition of planet building blocks (PBB) in the different galactic populations, using the elemental abundances determined with the HARPS survey \citep{Adibekyan2012b}. Using a simple stoichiometric model, they determined the molecular abundances and their respective mass fractions expected after the condensation sequence. Stellar synthetic models are particularly suitable for this kind of analysis allowing to study in details the properties of the thin and thick disks, halo and bulge. In line with the previous work of S17, we explore statistical trends of the PBB composition in the different galactic populations. We use the Besan\c con Galaxy Model (BGM) to simulate the global and chemical properties of stars in different populations \citep{Lagarde2017}. We then apply the S17's stoichiometric model to theses synthetic populations. This allows us to give robust predictions of the PBB for the whole Milky Way, not only the solar neighborhood. The development of such an coherent and integrated model is important to prepare the interpretation of future large scale surveys of exoplanets search.

The paper is structured as follows. In Sect. \ref{SecNumMet} we present the numerical tools we used in this work: the BGM model and the analytical prescriptions used to estimate the PBB composition. In Sect. \ref{SecMW1kpc} we simulate the solar neighborhood, and we compare our results with the values of mass fractions obtained by S17 with the HARPS sample data. In Sect. \ref{SecMW50kpc} we present our results for the overall Milky Way up to 50 kpc galactic distances from the Sun and the perspectives.

\section{Numerical method}
\label{SecNumMet}

\subsection{The Besançon Galaxy Model}
\label{SecBGM}

The Besan\c con stellar populations synthesis model provides the global (e.g., M, R, Teff) and chemical properties of stars for 54 chemical species. To reproduce the overall galaxy formation and evolution, four populations are considered: the halo, the bulge, the thin and thick disks. Each assumes different initial mass functions (IMF) and different formation and evolution histories. In each case, the IMF is a three-slopes power law. In the thin disk population the star formation rate is assumed decreasing exponentially with time following \citet{AumerBinney2009}. The parameters of the IMF and star formation history (SFH) of the thin disk have been fit to the Tycho-2 catalog \citep{Czekaj2014}. The IMF and SFH of the thick disk and halo have been set from comparisons to photometric data from SDSS and 2MASS surveys \citep[][]{Robin2014}.

As presented by \citet[][]{Lagarde2017} and \citet[][]{Lagarde2018}  (see for more details), the new version of the BGM includes a new grid of stellar evolution models computed with the stellar evolution code STAREVOL \citep[e.g.][]{Lagarde2012, Amard2016} for stars with $M \ge$ 0.7 M$_\odot$. These stellar evolution tracks have been computed from the pre-main sequence to the early asymptotic giant branch at six metallicities ([Fe/H]= 0.51, 0, -0.23, -0.54, -1.2, -2.14), and at different  $\alpha$-enhancements ([$\alpha$/Fe]=0.0, 0.15 and 0.30) to simulate all populations.

The data release 12 of the APOGEE spectroscopic survey \citep{Majewski2017} has been used to determine the [$\alpha$/Fe]-[Fe/H] trend, for the four galactic populations: 

\begin{equation}
      \mbox{[$\alpha$/Fe]}=
     \left\{
     \begin{array}{lll}
     0.014+0.01406\times\mbox{[Fe/H]} \\
\hspace{0.9cm}  +0.1013\times\mbox{[Fe/H]}^{2} & \mbox{Thin disk, [Fe/H]<0.1} \\
    0 & \mbox{Thin disk, [Fe/H]>0.1}\\
     0.320-\mbox{e}^{(1.19375\times\rm{[Fe/H]}-1.6)} & \mbox{Thick disk and bulge} \\
   0.3 & \mbox{Halo} \\
     \end{array}
     \right.
\end{equation}

\begin{figure}[h]
  \centering
    \includegraphics[width=\hsize,clip=true,trim= 0.1cm 0.1cm 0.1cm 0.1cm]{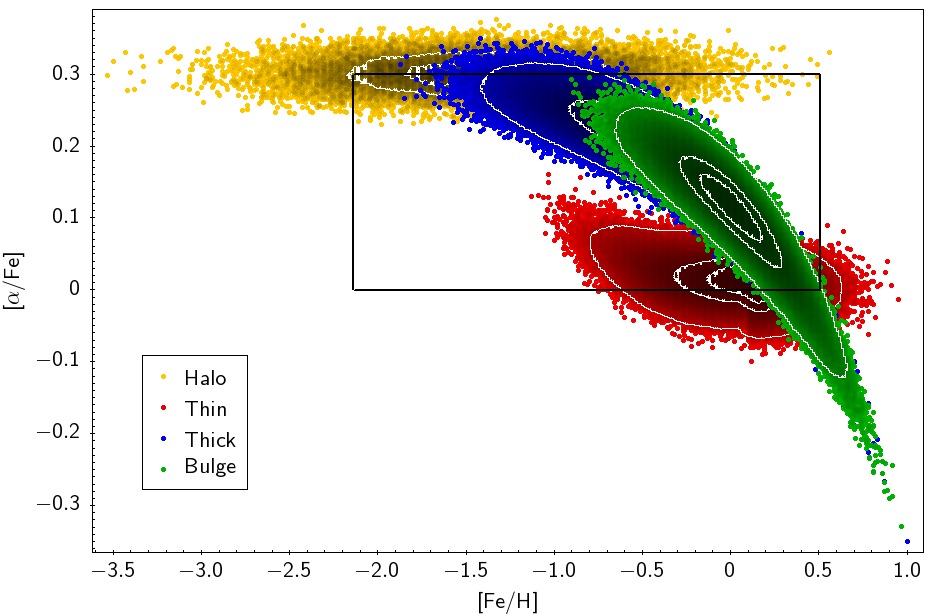}
\caption{The [$\alpha$/Fe] abundance as a function of [Fe/H] for stars with d<50 kpc simulated with the BGM. Thin and thick disks as well as bulge and halo are represented by red, blue, green and yellow dots respectively. Black lines indicate the selected population used in this study: -2.14<[Fe/H]<0.51 and 0<[$\alpha$/Fe]<0.3.}
    \label{FigAlpha}
\end{figure}

To these relations an intrinsic Gaussian dispersion of 0.02 dex is added. Although this improvement provides a better simulation of all stellar populations in our Galaxy, we are limited to the border of the stellar grid in metallicity and alpha content -2.14<[Fe/H]<0.51 and 0<[$\alpha$/Fe]<0.3 (inside black lines on Fig.\ref{FigAlpha}). These limits will be removed in the future as new stellar models are computed. Still 81\% of the whole sample remain after these cuts. We note that so far there are no planets detected outside our stellar metallicity limits -2.14<[Fe/H]<0.51. Even if future detections are not excluded, theoretical works derived a critical metallicity to form planets \citep[see e.g.][]{JohnsonLi2012}. Moreover, the vast majority of the observed planet host star are in the 0<[$\alpha$/Fe]<0.3 limits or very close \citep[see e.g.][]{Adibekyan2011}.

\subsection{Stoichiometric model}
\label{SubStoichio}

Elemental abundance ratios, as C/O and Mg/Si\footnote{We remind that A/B$\neq$[A/B]. We use, A/B=N$_A$/N$_B$=10$^{log\epsilon(A)}$/10$^{log\epsilon(B)}$, where $log\epsilon(A)$ and $log\epsilon(B)$ are the absolute abundances.}, govern the distribution and formation of chemical species in the protoplanetary disk. Because stellar atmospheres evolve slowly, the composition of a protoplanetary disk (in early evolutive phases) can be assumed to be the same as the star composition \citep{Lodders2003, Bond2010}. When solids condense from the gaseous disk, these ratios determine the planetesimal geology \citep{Pontoppidan2014}. If the C/O ratio is above $\sim$0.8, there is almost no free oxygen available to form silicates. Then the geology of planetesimals will be dominated by carbonates \citep{KuchnerSeager2005,Bond2010}. When the C/O ratio is below $\sim$0.8, planetesimals are dominated by magnesium bearing silicates. In these systems, the silicate distribution will be determined by the local Mg/Si ratio. There are three regimes of mineral formation: 

(1) when Mg/Si<1, the magnesium forms primarily pyroxene (MgSiO$_{3}$) and the remainder of the silicon forms feldspars or olivine (Mg$_{2}$SiO$_{4}$). 

(2) when 1<Mg/Si<2 there is an equally distributed mixture of pyroxene and olivine  similar to the solar system. 

(3) when Mg/Si>2, silicon forms olivine and the remainder of magnesium will forms magnesium compounds such as MgO and MgS.

We use the stoichiometric model published in \citet{Santos2017}. In this simple model, the molecular abundances in the protoplanetary disk, and their mass fraction, can be computed from the stellar abundances of a handful of elements. Fe, Si, Mg, O and C together with H and He control the species expected from the equilibrium condensation of H$_{2}$, He, CH$_{4}$, H$_{2}$O, Fe, MgSiO$_{3}$, Mg$_{2}$SiO$_{4}$ and SiO$_{2}$ \citep{Lodders2003, Bond2010}. These compounds dominate the rocky interior of Earth-like planet \citep[see e.g.][]{Sotin2007}.

Since we limited our synthetic population to the borders of the stellar grids (-2.14<[Fe/H]<0.51, 0<[$\alpha$/Fe]<0.3) all simulated stars have 1<Mg/Si<2. We write here the inverted stoichiometric relations corresponding to the case 1<Mg/Si<2 (assuming the equations on Appendix B of S17):

\begin{equation}
\begin{split}
 & N_{MgSiO_{3}} = 2 N_{Si} - N_{Mg} \\
 & N_{Mg_{2}SiO_{4}} = N_{Mg} - N_{Si} \\
 & N_{H_{2}O} = N_{O} - 2 N_{Si} - N_{Mg} \\
 & N_{CH_{4}} = N_C 
\end{split}
\end{equation}

with N$_X$ the number of atoms of each specie $X$. 
These relations enable the computation of the expected mass fractions of PBB : the iron-to-silicate mass fraction ($f_\text{iron}$), the water mass fraction ($f_\text{w}$) and the summed mass percent of all heavy elements ($f_\text{Z}$). The expression of these fractions are given in equation (1), (2) and (3) of S17.

\section{Comparison of the solar neighborhood simulation with the HARPS survey} 
\label{SecMW1kpc}

The PPB compositions have been determined using the method described in section \ref{SubStoichio}. Using the BGM, we simulate the close solar neighborhood, up to 100 pc (hereafter called \textit{HARPS simulation}), and the solar neighborhood, up to 1 kpc (hereafter called \textit{SN simulation}). We focus on FGK stars since they directly compare to the work of S17 on the HARPS survey. They are obviously interesting targets for the future exoplanetary surveys as TESS, PLATO and CHEOPS. Moreover, most of the current detected planets belong to the solar neighborhood \citep[from the database available at the Extrasolar Planets Encyclopaedia\footnote{http://exoplanet.eu},][ the large majority of exoplanets are within 1 kpc]{Schneider2011}, it is then interesting to study the PBB chemical composition at close distances from the Sun.

\subsection{HARPS simulation}

The target selection of the HARPS sample can not be simulated by a clear selection function and prevents us to model rigorously a synthetic HARPS population. In the following, we present a qualitative comparison between the results obtained by S17 with the HARPS survey and our results obtained with the \textit{HARPS simulation}. When restricting our synthetic population to the heliocentric distance, the metallicity range and the effective temperature range of the HARPS sample used in S17, d<100 pc, -0.83<[Fe/H]<0.41 and stars with an effective temperature of +-300K around solar T$_{\rm{eff},\odot}$=5777K  (see S17), we have 10237 stars: 87\% of thin disk stars and 13\% of thick disk stars. Instead, the HARPS sample used by S17 includes 303 (81.7\%) thin disk stars and 68 (18.3\%) thick disk stars. We stress that the BGM can simulate all stars satisfying the mentioned range in metallicity, effective temperature and galactic distance. Instead, the HARPS sample includes only a group of them, introducing potentially a bias which can not be take into account. Since the selection function of the HARPS survey can not be derived, the consequent observational bias could introduce differences between the model and the results obtained with observations.

\begin{figure}
  \centering
    \includegraphics[width=9cm,clip=true,trim= 0cm 0cm 0cm 0cm]{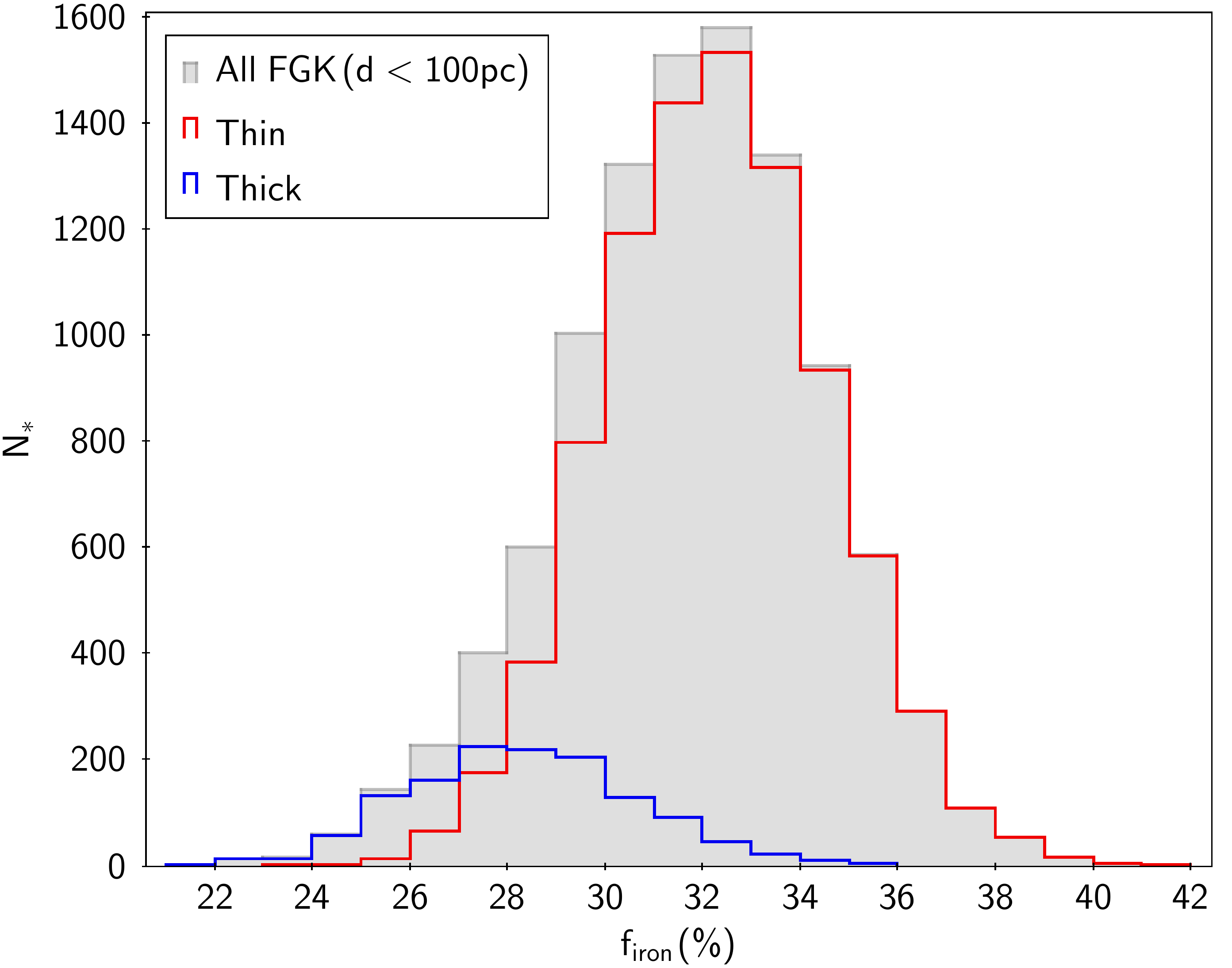}
    \vskip 1cm
    \includegraphics[width=9cm,clip=true,trim= 0cm 0cm 0cm 0cm]{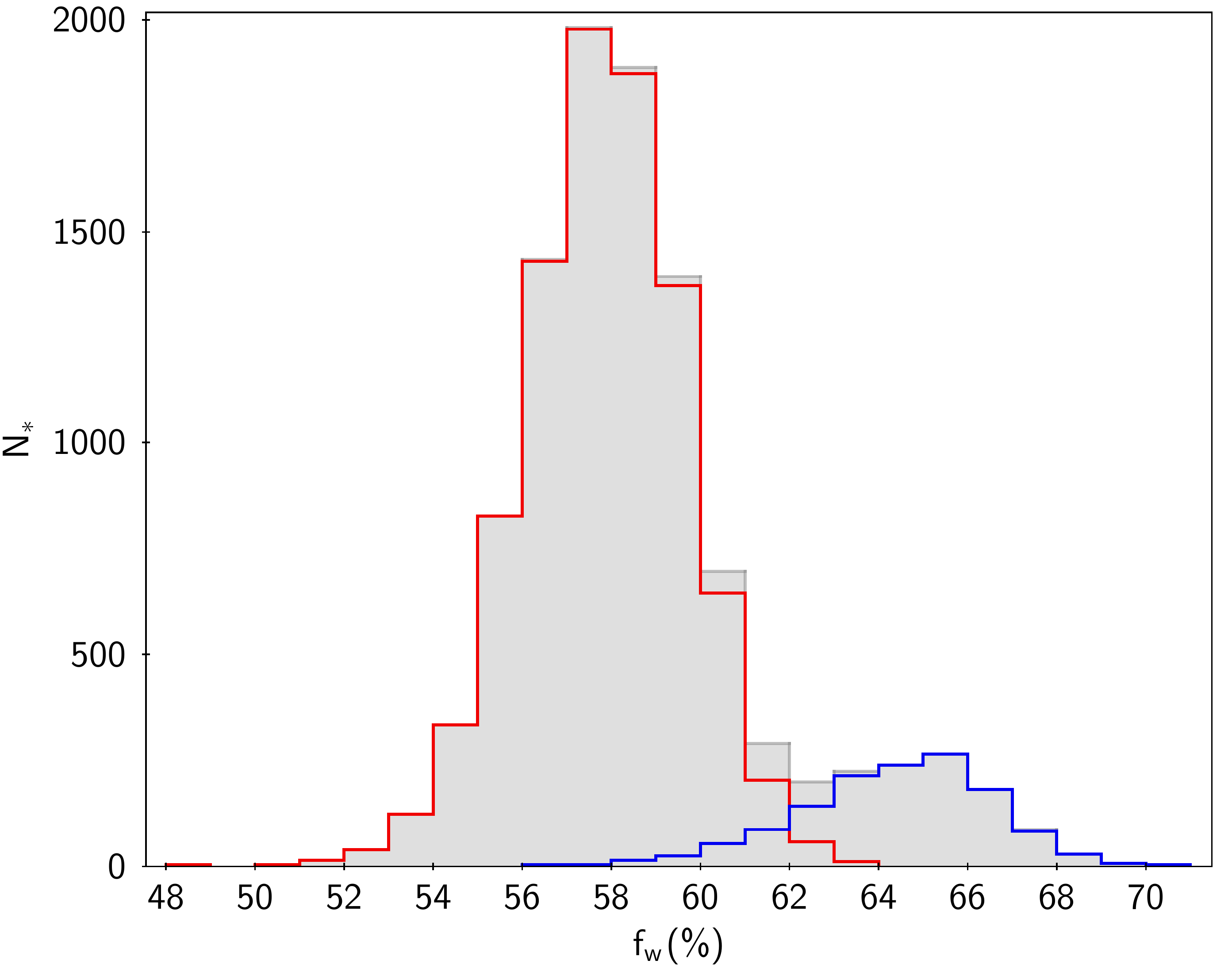}
    \vskip 1cm
    \includegraphics[width=9cm,clip=true,trim= 0cm 0cm 0cm 0cm]{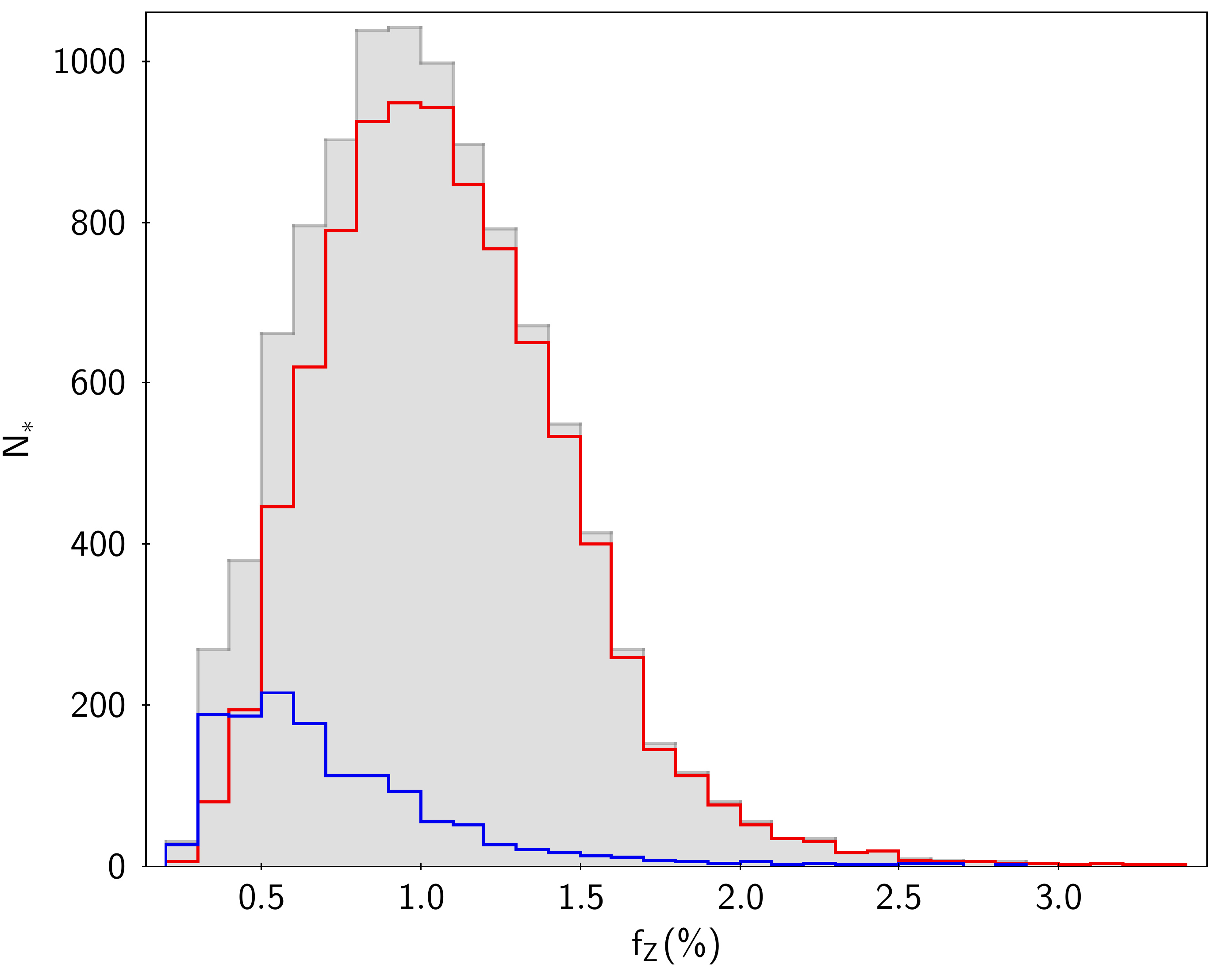}
\caption{Mass fraction distributions, $f_\text{iron}$ (upper panel), $f_\text{w}$ (middle panel) and $f_\text{Z}$ (lower panel) for the three stellar populations of the Milky Way: thin disk, thick disk and halo. We ran the model up to distances of 100 pc.}
    \label{Fig_d1}
\end{figure}

Moreover, the HARPS sample has intrinsic errors on the observed chemical abundances. To compare with the HARPS observations, the observed mean errors \citep[see][]{Adibekyan2011,BertranDeLis2015,Suarez_Andres2017} are added to the \textit{HARPS simulation} using a gaussian dispersion. Figure \ref{Fig_d1} displays the synthetic distributions of the mass fraction of iron, water and heavy elements of the PBB around stars for each galactic stellar populations in the solar neighborhood.

\begin{table}
\centering                                      
\caption{Summary of the simulations content.}  
\begin{tabular}{lccc}          
\hline                      
                  & MW simul.      & SN simul.   & HARPS simul.\\    
\hline                                   
   d              & <50 kpc        &  <1 kpc     & <100 pc\\
   N$_*$          & 4 850 600 000 &  14 850 705 & 34 162\\
   Thin (\%)      & 42        &  78.99      & 90.81\\
   Thick (\%)     & 27        &  21         & 9.19\\
   Halo (\%)      & 1         &  <0.001     & <0.001 \\
   Bulge (\%)     & 30        &             &  \\
   \hline                      
\end{tabular}
\label{tableModels}
\end{table}

As shown in Figure \ref{Fig_d1} (upper panel), the majority of the synthetic thick disk stars have iron mass fractions within 21-36\%. Using the HARPS survey, S17 found that the stellar thick disk population has iron mass fractions ranging in the same interval. For the thin disk, synthetic $f_\text{iron}$ ranges from 26-38\% in the simulation and from $\sim$25-37\% in the HARPS sample.

\begin{figure}
  \centering
    \includegraphics[width=\hsize,clip=true,trim= 0cm 0cm 0cm 0cm]{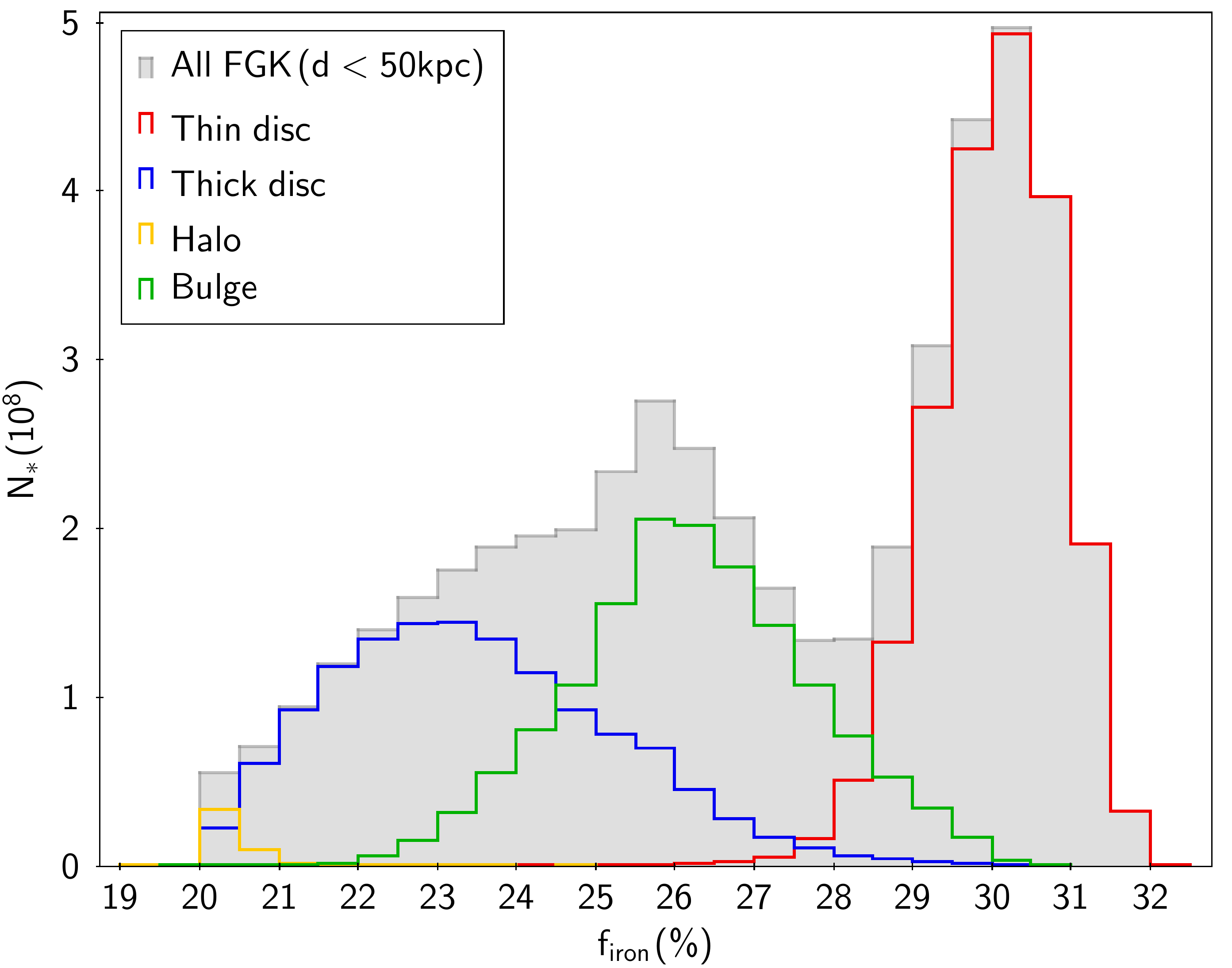}
    \vskip 1cm
    \includegraphics[width=\hsize,clip=true,trim= 0cm 0cm 0cm 0cm]{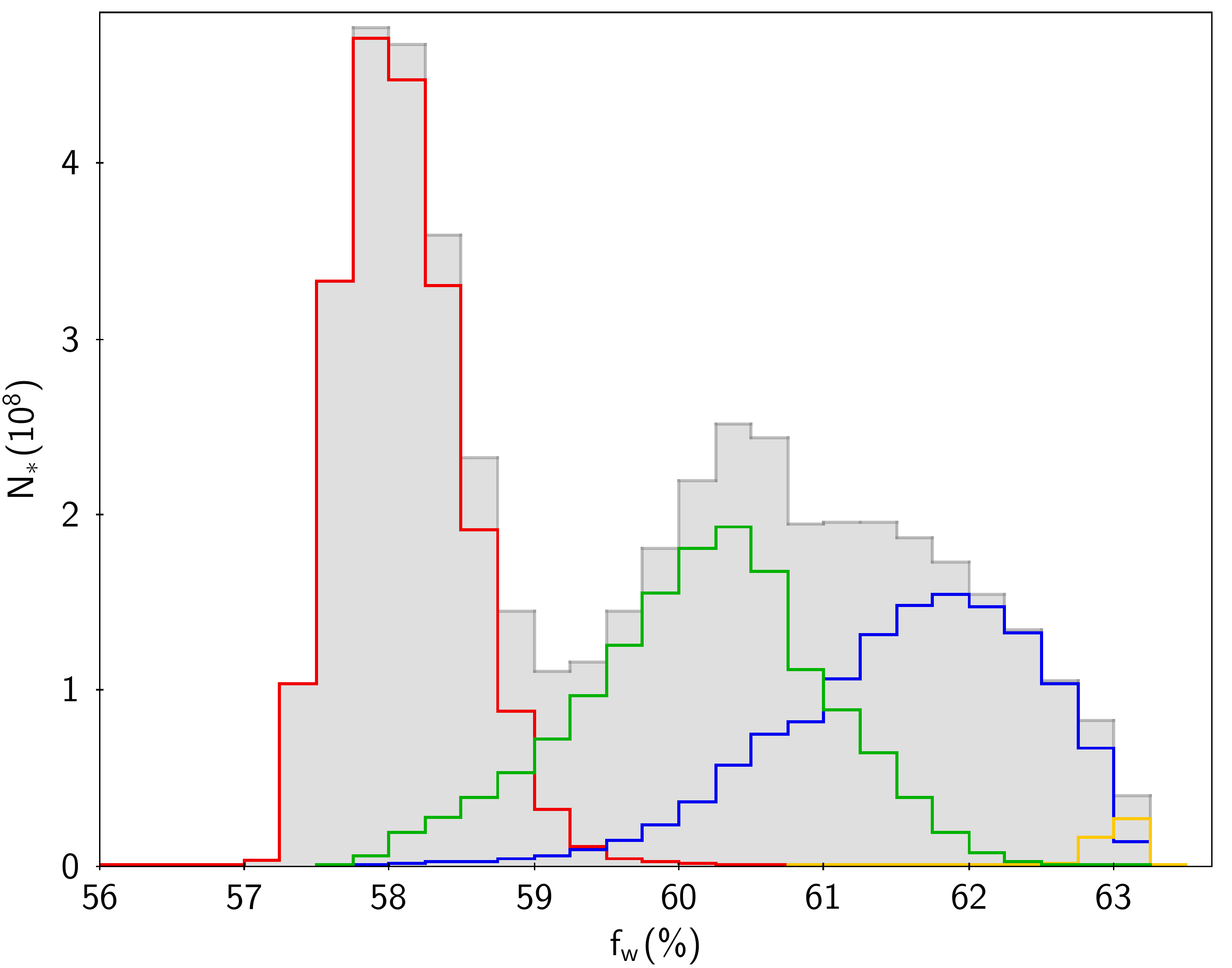}
    \vskip 1cm
	\includegraphics[width=\hsize,clip=true,trim= 0cm 0cm 0cm 0cm]{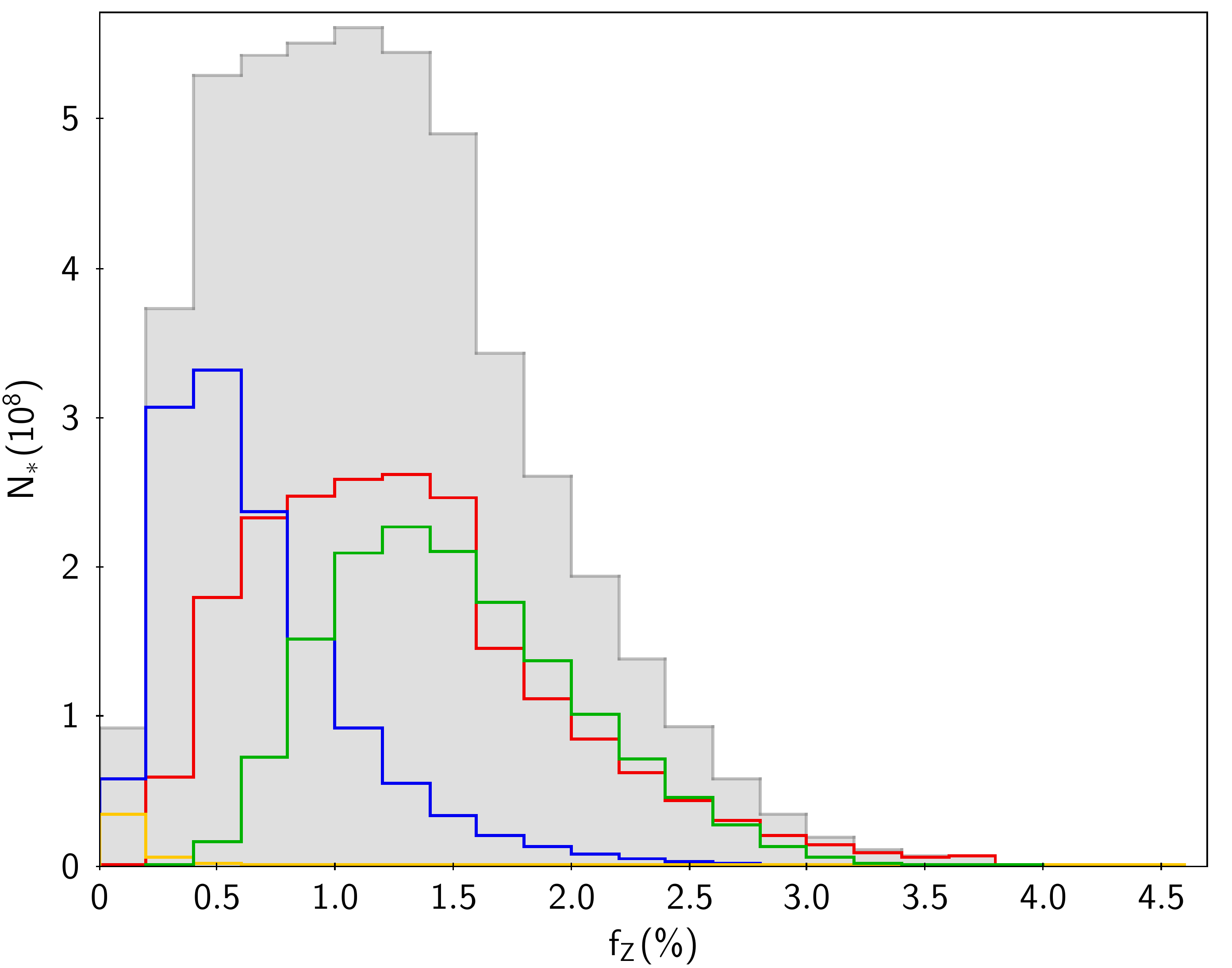}
     \caption{Mass fraction distributions, $f_\text{iron}$ (upper panel), $f_\text{w}$ (middle panel) and $f_\text{Z}$ (lower panel) for the four stellar populations of the Milky Way: thin disk, thick disk, halo and bulge. We ran the model up to distances of 50 kpc.}
    \label{FigMW50Histo}
\end{figure}

The synthetic thin disk water mass fraction ranges in $\sim$48-64\%, while S17 found $\sim$40-80\%. The synthetic thick disk ranges within $\sim$56-72\%, while S17 found $\sim$40-90\%. For the heavy elements $f_\text{Z}$, the synthetic thin disk ranges in $\sim$0-3.5\%, while S17 found $\sim$0.5-3.5\%. The synthetic thick disk ranges within $\sim$0-2.8\%, while S17 found $\sim$0.5-2.6\%. Overall we find good agreement between the distribution of water mass fraction in the simulated and the observed sample. However, we note that our synthetic intervals are less larger than those derived by S17 using HARPS data. This could be due to the limits of the stellar grid we use in the model, since we show in section \ref{SecMW50kpc} that $f_\text{w}$ is manly dependent on [$\alpha$/Fe] (see Fig. \ref{Fig_Alpha_firon}). Yet, the differences may be partly explained by the typical high errors in the derivation of observed abundances for Oxygen \citep[see e.g.][]{BertranDeLis2015,Suarez_Andres2017}, which enter in the calculation of $f_\text{w}$.

\subsection{Solar neighborhood simulation}

Table \ref{tableAverage} summarizes the average values of $f_\text{iron}$, $f_\text{w}$ and $f_\text{Z}$ and their corresponding standard deviations for PBB belonging to each stellar population. We list in the table the results obtained for the three simulations of this work: the \textit{HARPS simulation}, the \textit{SN simulation} and the \textit{Milky Way (MW) simulation} (discribed in next section). To be more general and not focus on specific survey, we do not include Gaussian dispersion on the chemical abudnances for the \textit{SN simulation} and the \textit{MW simulation}. In addition, the results for the \textit{HARPS simulation} are given with and without gaussian dispersion. 

The \textit{SN simulation}, includes larger galactic distances (d<1kpc) than the \textit{HARPS simulation} (d<100pc). Again to be more general, the \textit{SN simulation} is not restricted to the HARPS range in metallicity and in effective temperature.  The synthetic sample has a total of 14 850 705 stars: 78.9\% from the thin disk, 21\% from the thick disk and <0.001\% from the halo. Our simulation shows that the average values are very similar for d<100 pc (\textit{HARPS simulation} with and without gaussian dispersion) and for d<1 kpc (\textit{SN simulation}) within a given galactic population. However, we should keep in mind that the proportion of thin and thick disk stars are substantially different. For d<1 kpc the fraction of thick disk stars is double (21\%) than for d<100 pc (13\%). This impacts the average values of all galactic populations in Table \ref{tableAverage}. Indeed, when including a larger fraction of thick stars at our synthetic sample, we actually include more $\alpha$-enhanced stars, and so $f_\text{iron}$ decreases while $f_\text{w}$ decreases. The $\alpha$-enhanced stars are also less metallic and therefore $f_\text{Z}$ decreases. The average PBB composition is actually changing as function of the heliocentric distance.

In spite of the weak statistics available in the study of S17 (371 stars), we notice the overall good agreement with our simulations, in the iron, water and heavy element mass fractions. We developped a new version of the BGM to simulate the PBB chemical composition in the solar neighborhood using the stoichiometric model developped by S17 and taking into account stellar and galactic evolution. In the following section we extend the analysis to the whole Galaxy.

\begin{table*}
\centering                                      
\caption{The average mass fraction values for iron and water, and for the mass percent of heavy elements , in the different galactic populations. We also report values found by the stoichiometric model for the Sun. Values are expressed in percentage and their standard deviations are given.}  
\begin{tabular}{lccc}          
\hline                      
& <$f_\text{iron}$> & <$f_\text{w}$> &  <$f_\text{Z}$>\\    
\hline                      
 \textbf{HARPS simulation (d<100pc, including gaussian dispersion)}   &  &  &  \\    

   Thin            &  31.8 $\pm$ 2.6 &  58.7 $\pm$ 2.9  &  1.0 $\pm$ 0.4\\
   Thick           &  28.3 $\pm$ 2.3 &  64.4 $\pm$ 2.1  &  0.7 $\pm$ 0.4\\
   All populations &  31.8 $\pm$ 2.6 &  58.7 $\pm$ 2.9  &  1.0 $\pm$ 0.4\\
\hline                      
 \textbf{HARPS simulation (d<100pc)}   &  &  &  \\    
   Thin            &  30.3 $\pm$ 0.7 &  58.0 $\pm$ 0.4  &  1.1 $\pm$ 0.4\\
   Thick           &  23.6 $\pm$ 1.5 & 61.5 $\pm$ 0.7  &  0.7 $\pm$ 0.4\\
   All populations &  29.4 $\pm$ 2.4 &  58.5 $\pm$ 1.3  &  1.0 $\pm$ 0.4\\
\hline                      
\textbf{HARPS, S17}      & && \\    
   Thin   &    31.974 $\pm$ 1.750 &  59.713 $\pm$ 7.106  &  1.508 $\pm$ 0.597\\
   Thick  &    24.305 $\pm$ 1.623 &  72.179 $\pm$ 5.961  &  1.135 $\pm$ 0.295\\
   Halo$^{a}$           &    23.110 $\pm$ 2.884 &  83.990 $\pm$ 4.115  &  0.808 $\pm$ 0.082\\
\hline                      
\textbf{Sun} &    33           &  60            &  1.3          \\
\hline                      
\hline                      
 \textbf{SN simulation (d<1kpc)}   &  &  &  \\    
   Thin           &  30.2 $\pm$ 0.8 &  58.0 $\pm$ 0.4  &  1.1 $\pm$ 0.4\\
   Thick          &  23.5 $\pm$ 1.7 &  61.6 $\pm$ 0.9  &  0.7 $\pm$ 0.4\\
   Halo           &  20.4 $\pm$ 0.2 &  63.0 $\pm$ 0.1  &  0.2 $\pm$ 0.3\\ 
   All populations &  28.7 $\pm$ 2.9 &  58.8 $\pm$ 1.5  &  1.0 $\pm$ 0.4\\                                      

\hline                      %
\textbf{MW simulation (d<50kpc)}   & & &\\
   Thin           &    30.0 $\pm$ 0.8 &  58.1 $\pm$ 0.4  &  1.3 $\pm$ 0.6\\
   Thick          &    23.5 $\pm$ 1.7 &  61.6 $\pm$ 0.9  &  0.7 $\pm$ 0.4\\
   Halo           &    20.4 $\pm$ 0.2 &  63.0 $\pm$ 0.1  &  0.2 $\pm$ 0.2\\
   Bulge          &    26.2 $\pm$ 1.5 &  60.1 $\pm$ 0.8  &  1.5 $\pm$ 0.5\\
   All populations &    27.0 $\pm$ 3.1 &  59.7 $\pm$ 1.6  &  1.2 $\pm$ 0.6\\                                      
\hline                      
\multicolumn{3}{l}{$a$ - Only three halo stars are included in this galactic population.}
\end{tabular}

\label{tableAverage}
\end{table*}

\section{Chemical trends in the Milky Way predicted by the BGM}
\label{SecMW50kpc}

We now investigate the general trends for the whole Milky Way. We thus simulate the FGK sample to distances up to 50 kpc from the Sun (hereafter called \textit{MW simulation}). This large volume covers the Galaxy up to the external parts. As for the \textit{SN simulation}, we do not include Gaussian dispersion on the chemical abundances for this simulation, to be general and to avoid to focus on a specific survey. Our simulation has a total of 4 850 600 000 stars: 42\% from the thin disk, 27\% from the thick disk, 30\% from the bulge, and 1\% from the halo. 

Average values of $f_\text{iron}$, $f_\text{w}$ and $f_\text{Z}$ are given in Table \ref{tableAverage}. Figure \ref{FigMW50Histo} presents the distributions of iron, water and heavy elements mass fractions of PBB as a function of the galactic populations. The four populations present significant differences in their values of iron and water mass fractions. This means we can expect different compositions for rocky planets. 

All stellar populations taken together, the iron mass fraction is between 20\% and 32\%. The thin disk stars host PBB with higher iron mass fraction ($<f_\text{iron}>$= 30\%) than other stellar populations: $<f_\text{iron}>$= 23.5, 20.4, and 26.2\% for thick disk, halo, and bulge, respectively (see top panel of Fig. \ref{FigMW50Histo} and table \ref{tableAverage}). The water mass fraction of PBB in the thin disk ($<w_\text{f}>$= 58.1\%) is lower than other populations ($<w_\text{f}>$= 61.6, 63, and 60.1\% for thick disk, halo, and bulge, respectively, see middle panel of Fig. \ref{FigMW50Histo}). 

If we are able to distinguish thin disk stars from stars of the other populations we can deduce from our galactic model the approximate composition (iron and water mass fractions) of the planets observed by a given survey. Actually, determining the galactic population (using kinematics, e.g. from \textit{Gaia}, and/or accurate $\alpha$-abundances determination) could improve our knowledge on initial conditions, as planetesimal composition, of planet formation models.

\subsection{The water/iron valley}
\label{wivalley}

\begin{figure}
  \centering
    \includegraphics[width=\hsize,clip=true,trim= 0cm 2cm 0cm 0cm]{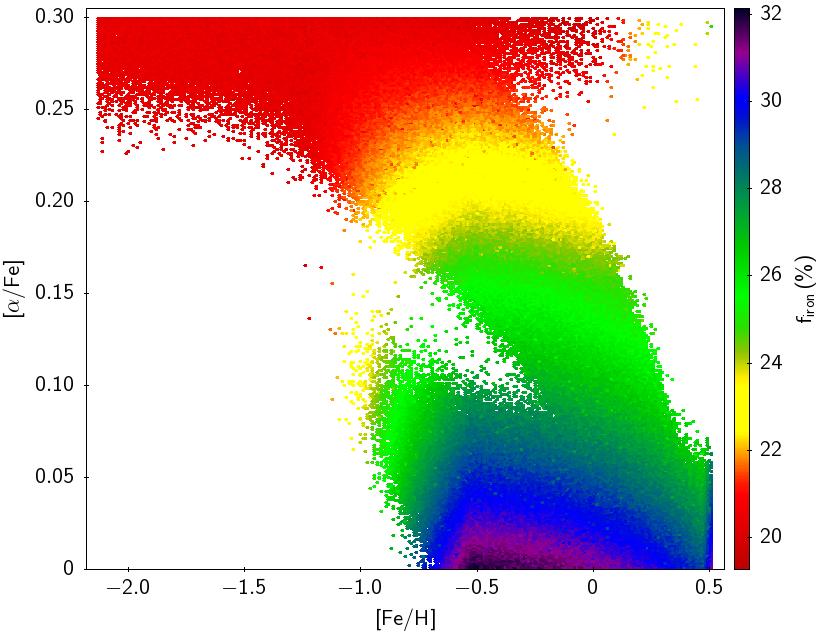}
    \includegraphics[width=\hsize,clip=true,trim= 0cm 2cm 0cm 0cm]{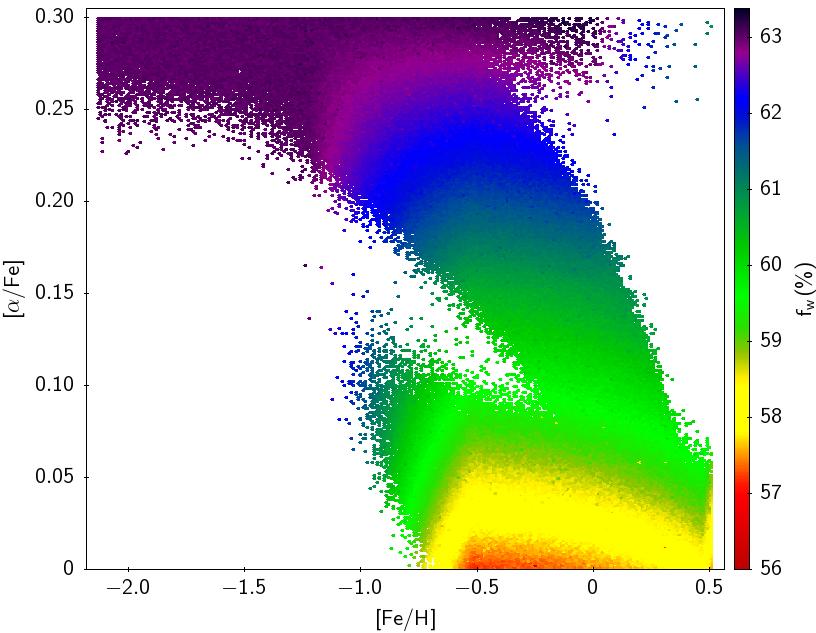}
	\includegraphics[width=\hsize,clip=true,trim= 0cm 0cm 0.078cm 0cm]{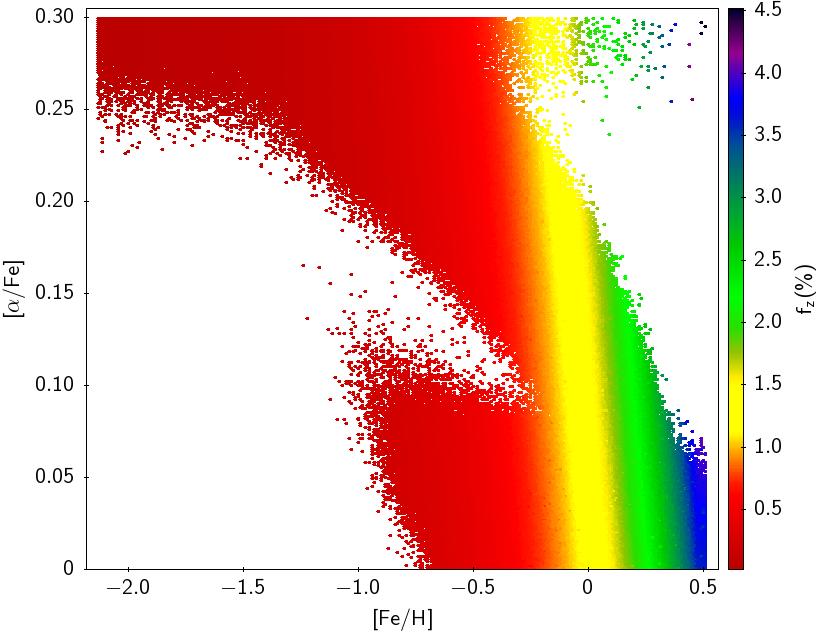}
\caption{The iron-to-silicate mass fraction, $f_\text{iron}$ (upper panel), the water mass fraction, $f_\text{w}$ (middle panel) and the heavy elements mass fraction $f_\text{Z}$ (lower panel) for the four stellar populations of the Milky Way: thin disk, thick disk, halo and bulge. We ran the model up to distances of 50 kpc.}
    \label{Fig_Alpha_firon}
\end{figure}

Figure \ref{Fig_Alpha_firon} shows the [$\alpha$/Fe] ratio of FGK stars simulated with the BGM as a function of stellar metallicity, color-coded  with the mass fraction of iron (top panel), water (middle panel) and heavy elements (bottom panel) of PBB. The stellar alpha abundance appears to be crucial when determining the iron and the water mass fractions of PBB. Indeed, Fig. \ref{Fig_Alpha_firon} shows that the iron (upper panel) and water (middle panel) fractions in PBB are strongly dependent on the alpha abundance in stars and less on their metallicity, whereas the proportion of heavy elements (bottom panel) essentially depends on the metallicity of the star. It is worth noting that S17 describe quantitatively a dependence of $f_\text{iron}$ with [Si/Fe] and qualitatively a dependance with [Fe/H] based on the chemical abundances observed by HARPS (371 stars). In the present work, using the BGM we show that $f_\text{iron}$ and $f_\text{w}$ are mainly function of [$\alpha$/Fe] (Fig.\ref{Fig_Alpha_firon}). We remind that the PBB compositions are dependent on the stellar abundances, which are here predicted by the evolutionary model STAREVOL. The detailed predictions of abundances with this model allows us to clearly show that $f_\text{iron}$ and $f_\text{w}$ are essentially dependent on alpha elements.

As seen in Fig. \ref{FigMW50Histo}, the synthetic population computed with the BGM predicts a distinct distribution between the thin disk stars with iron-rich PBB, and other stellar populations with water-rich PBB, implying a significant dip in the number of stars around $f_\text{iron}$$\sim$28\% and $f_\text{w}$$\sim$59\%. This "iron/water valley'' results from the stellar alpha content distributions  in the synthetic stellar populations. Indeed, the density of stars around solar metallicity and [$\alpha$/Fe]$\sim$0.1 is smaller due to the known gap between the thin and the thick disk. Since we derive that $f_\text{iron}$ and $f_\text{w}$ are mainly function of [$\alpha$/Fe] we understand why this gap translates into a bimodal distribution in the $f_\text{iron}$ and $f_\text{w}$ histograms. In other words, the synthetic population computed with the BGM shows the presence of the "iron/water valley'' because of the clear dependence of $f_\text{iron}$ and $f_\text{w}$ on [$\alpha$/Fe]. We underline that this valley is not dependent on the limits in metallicity (-2.14<[Fe/H]<0.51) and alpha abundances(0<[$\alpha$/Fe]<0.3) we use in this analysis.

Interestingly, the \textit{HARPS simulation} presents also a water valley around $f_\text{w}$=62\% (Fig. \ref{Fig_d1}) separating the thin and the thick disks. Moreover, $f_\text{iron}$ histograms from the thin and the thick disks intersect around 28\% as in the valley present in the MW simulation. The HARPS data is compatible with the iron/water valley (see the left panels of Fig. 1 in S17), albeit less clear that in our simulations due to their weak statistics (371 stars). This valley, not mentioned by S17, is clearly predicted by our simulation thanks to the larger statistics. We believe that better statistics, in particular larger spectroscopic surveys on FGK stars, may indirectly constrain the iron/water valley at any distances from the Sun. Moreover, if we are able to determine the galactic origin of stars in a larger sample, we could confirm our findings on the valley between the thin disc and the others galactic populations.

\subsection{Expected planet formation efficiency}

Observations show a clear correlation between giant planets frequency and the stellar metallicity. This is usually explained in the frame of the core accretion model \citep{Pollack1996}. Assuming that the protoplanetary disk metallicity scales with the stellar metallicity \citep{Pontoppidan2014}, then the amount of condensible solids is higher in metal-rich stars, and more building blocks are available for planet formation. In the context of the core accretion model, more building blocks implies a higher probability of collision \citep{Inaba2001} and consequently an enhanced efficiency for planet formation \citep[see e.g.][]{IdaLin2004}. Thus, the host star metallicity is believed to determine the amount of solid material available to be accreted by the planetary embryos.

As shown in bottom panel of Fig. \ref{Fig_Alpha_firon}, the heavy elements mass fraction $f_\text{Z}$ correlates with the stellar metallicity and weakly with the initial alpha abundances. Thus, the more-metallic stars of the thin disk and the bulge present higher average values of $f_\text{Z}$, 1.3\% and 1.5\% respectively, while metal-poor stars of the halo and the thick disk have average values of $f_\text{Z}$, 0.2\% and 0.7\% respectively. Values are summarized in Table \ref{tableAverage}. In the context of the core accretion mechanism, the bulge and the halo may have the largest, respectively lowest, frequency of giant planets of the Galaxy.

\subsection{Expected exoplanets composition}

From the expected chemical composition of planet building blocks we can discuss the expected exoplanets composition. As mentioned by S17, the stoichiometric model predicts for the Sun $f_\text{iron}$$\sim $$3\%$, $f_\text{w}$$\sim$$60\%$ and $f_\text{Z}$$\sim$$1.3\%$. This value of $f_\text{iron}$ is similar to the values observed in the meteorites and the rocky planets of the solar system, Earth, Venus and Mars \citep[see e.g.][]{DrakeRighter2002}. Actually, solar $f_\text{w}$ and $f_\text{Z}$ are also compatible with the values found by \citet{Lodders2003} ($f_\text{w}$=67.11\% and $f_\text{Z}$=1.31\%).

When we compare the solar values with the values obtained by the \textit{HARPS simulation} and the \textit{SN simulation}, we observe that the synthetic thin disk stars present PBB chemical composition compatible with the values of the solar system. Actually, stars of the thin disk could produce planets with similar composition to the rocky planets of the solar system. Metal-poor stars of the thick disk might produce planets with lower iron mass. Similar results are obtained by S17 with HARPS observations.

Additionally, we derive in Sect.\ref{SecMW1kpc} that the PBB composition is function of the heliocentric distance in the solar neighborhood. This is because the fraction of each galactic population is changing with the heliocentric distance. The \textit{MW simulation} includes naturally less thin disk stars (42\% in the \textit{MW simulation} and $\sim$79\% in the \textit{SN simulation}), such that the fraction of rocky planets may decrease. On the other hand, the bulge presents the closest PBB compositions to the solar ones, and should also be able to host rocky planets with solar-system-like composition. Halo stars should produce planets with low iron mass fraction but rich water content. Currently there is only one planetary mass object detected in the bulge, the brown dwarf OGLE-2017-BLG-1522Lb \citep{Jung2018} and there is no planet detected orbiting halo stars.

\section{Conclusion}

We compare the chemical composition trends of planet building blocks in the different galactic populations of the Milky Way. To this aim, we compute stellar population synthesis including the simulation of stellar surface chemical abundances. Then, we use a simple stoichiometric model derived by S17 to estimate the planet building blocks composition from the chemical properties of stars generated by the BGM. This work extends the study of S17 with HARPS data to the whole Galaxy. Our results are obtained with stellar chemical abundances predicted by the BGM and not from observations, which makes comparisons particularly interesting. The new version of the BGM, including the simple stoichiometric model derived by S17, appears as a powerful tool to predict chemical composition of PBB, crucial to prepare interpretation of on-going and future large scale surveys of exoplanet search.

We ran the BGM to generate a synthetic sample of FGK stars. In the simulation of the Milky Way we compute the model up to 50 kpc, to establish general trends for our Galaxy. In the simulation of the solar neighborhood, we analyse trends up to 1 kpc, where the majority of exoplanets have been observed. We also compare the S17 results obtained with the HARPS survey. Overall, we found a very good agreement with the results of S17 based on the HARPS observations. In spite of the weak statistics of the HARPS sample, the range of values on $f_\text{iron}$, $f_\text{w}$ and $f_\text{Z}$ are similar.

We list here the main results obtained with the Milky Way simulation:

\begin{itemize}  

\item \textsf{Alpha content dependence:} Our simulations show a clear dependence of both, iron and water mass fractions, with the initial alpha content [$\alpha$/Fe] of the host stars. We have shown that this dependence explains well the iron/water valley. Mass fractions are found to be also dependent with the metallicity but to a lesser degree. These dependencies have strong implications on the chemical composition trends of PBB with the galactic populations.\\

\item \textsf{Iron/water valley:} The different galactic populations (thin and thick disc, halo and bulge) are known to have different metallicities and alpha content [$\alpha$/Fe] distribution. The clear dependence on [$\alpha$/Fe] leads to a bimodal distribution of $f_\text{iron}$ and $f_\text{w}$ histograms between the thin disk stars and the other galactic populations: the iron/water valley. The valley is predicted by our simulations to be at $f_\text{iron}$$\sim$$28\%$ and $f_\text{w}$$\sim$$59\%$. \\

\item \textsf{PBB trends:} Our solar neighborhood simulations (d<1 kpc) shows that the thin disk is expected to present iron rich and water poor PBB, while the thick disk should have iron poor but water rich PBB. Similar results are found by S17 with the HARPS observations, although limited to 100 pc. In this work we simulate the whole Milky Way enabling to study the bulge population as well. The bulge appears to have intermediate values of $f_\text{iron}$ and $f_\text{w}$, between the ones of the thin and thick discs. Its mass fraction values overlap with the ones of the thin disc. It may produce rocky planets similar to the ones of the solar system. We definitively found a dependence of the PBB composition with the heliocentric distance. Indeed, the fraction of thin disk stars is decreasing with distance while the thick disk and bulge stars are increasing. These latter are predicted to be alpha-rich which impacts the values of $f_\text{iron}$ and $f_\text{w}$.   \\ 

\end{itemize}

It is important to mention a limitation related to the determination of the galactic origin of observed stars. To rigorously determine the galactic population of an observed star is currently not a straightforward task. There is no irrevocable criteria (or combination of criteria) that can give with certainty whether it belongs the thin or the thick disc. The galactic populations origin of the HARPS sample have been determined following criteria based on the alpha-content and the metallicity \cite[see][]{Adibekyan2012b}. A modified classification method for the observed HARPS populations could potentially lead to different PBB composition trends. Chemical abundances from spectroscopic data combined to the precise astrometry of the Gaia mission \citep{GaiaCollPrusti2016}, will allow to characterize better the galactic populations and consequently improve considerably the comparative analyses as the one presented in this work. Cross matching planet properties from the future large exoplanetary surveys and the galactic origin of the host stars based on the Gaia data should provide precise constraints on, for instance, the iron/water valley.

Several studies showed that the bulk composition is crucial to determine the planet internal structure. Typically, when planetary radius and mass are known, the mass and radius of the different internal layers remain largely degenerated \citep{Valencia2007}. It has been shown that the chemical bulk composition (in particular Mg/Si and Fe/Si ratios) enables to fix the detailed internal structure \citep[see e.g.][]{Dorn2015}. Thus it becomes crucial to estimate the chemical bulk composition. As soon as we keep in mind that the calculations done in this work are not a prediction of the final planet composition, the present paper provides some answers. Indeed, the mass fractions computed here should be used as indicators of tendencies for PBB composition and as proxies for expected planet chemical composition. The present results may be used as a starting step to predict the expected rocky planets properties in the different galactic populations. 

The determinations of accurate stellar and planetary masses and radii distributions from TESS, CHEOPS and PLATO space-missions will provide unique constrains to the current predicted statistical trends of planet composition (or proxy of those compositions as PBB compositions). Moreover, we should mention that, so far, the stellar grids of the BGM, include stellar abundances only for $M$>0.7 $M_\odot$. This excludes the large majority of M stars. The latter are thought to be interesting target to discover rocky planets in the habitable zones, and actually one of the main goal of the TESS mission, as well as ground-based large programs with HARPS \citep{Bonfils2013} and the new instrument SPIRou at Canada France Hawaii Telescope \citep{Delfosse2013}. The present study will be extended to M stars in a future work.

\bibliographystyle{aa}
\bibliography{reference}

\begin{thebibliography}{50}
\expandafter\ifx\csname natexlab\endcsname\relax\def\natexlab#1{#1}\fi

\bibitem[{{Adibekyan} {et~al.}(2015){Adibekyan}, {Santos}, {Figueira}, {Dorn},
  {Sousa}, {Delgado-Mena}, {Israelian}, {Hakobyan}, \&
  {Mordasini}}]{Adibekyan2015}
{Adibekyan}, V., {Santos}, N.~C., {Figueira}, P., {et~al.} 2015, \aap, 581, L2

\bibitem[{{Adibekyan} {et~al.}(2013){Adibekyan}, {Figueira}, {Santos},
  {Mortier}, {Mordasini}, {Delgado Mena}, {Sousa}, {Correia}, {Israelian}, \&
  {Oshagh}}]{Adibekyan2013}
{Adibekyan}, V.~Z., {Figueira}, P., {Santos}, N.~C., {et~al.} 2013, \aap, 560,
  A51

\bibitem[{{Adibekyan} {et~al.}(2011){Adibekyan}, {Santos}, {Sousa}, \&
  {Israelian}}]{Adibekyan2011}
{Adibekyan}, V.~Z., {Santos}, N.~C., {Sousa}, S.~G., \& {Israelian}, G. 2011,
  \aap, 535, L11

\bibitem[{{Adibekyan} {et~al.}(2012{\natexlab{a}}){Adibekyan}, {Santos},
  {Sousa}, {Israelian}, {Delgado Mena}, {Gonz{\'a}lez Hern{\'a}ndez}, {Mayor},
  {Lovis}, \& {Udry}}]{Adibekyan2012a}
{Adibekyan}, V.~Z., {Santos}, N.~C., {Sousa}, S.~G., {et~al.}
  2012{\natexlab{a}}, \aap, 543, A89

\bibitem[{{Adibekyan} {et~al.}(2012{\natexlab{b}}){Adibekyan}, {Sousa},
  {Santos}, {Delgado Mena}, {Gonz{\'a}lez Hern{\'a}ndez}, {Israelian}, {Mayor},
  \& {Khachatryan}}]{Adibekyan2012b}
{Adibekyan}, V.~Z., {Sousa}, S.~G., {Santos}, N.~C., {et~al.}
  2012{\natexlab{b}}, \aap, 545, A32

\bibitem[{{Amard} {et~al.}(2016){Amard}, {Palacios}, {Charbonnel}, {Gallet}, \&
  {Bouvier}}]{Amard2016}
{Amard}, L., {Palacios}, A., {Charbonnel}, C., {Gallet}, F., \& {Bouvier}, J.
  2016, \aap, 587, A105

\bibitem[{{Aumer} \& {Binney}(2009)}]{AumerBinney2009}
{Aumer}, M. \& {Binney}, J.~J. 2009, \mnras, 397, 1286

\bibitem[{{Beaug{\'e}} \& {Nesvorn{\'y}}(2013)}]{BeaugeNesvorny2013}
{Beaug{\'e}}, C. \& {Nesvorn{\'y}}, D. 2013, \apj, 763, 12

\bibitem[{{Bertran de Lis} {et~al.}(2015){Bertran de Lis}, {Delgado Mena},
  {Adibekyan}, {Santos}, \& {Sousa}}]{BertranDeLis2015}
{Bertran de Lis}, S., {Delgado Mena}, E., {Adibekyan}, V.~Z., {Santos}, N.~C.,
  \& {Sousa}, S.~G. 2015, \aap, 576, A89

\bibitem[{{Bond} {et~al.}(2010){Bond}, {O'Brien}, \& {Lauretta}}]{Bond2010}
{Bond}, J.~C., {O'Brien}, D.~P., \& {Lauretta}, D.~S. 2010, \apj, 715, 1050

\bibitem[{{Bonfils} {et~al.}(2013){Bonfils}, {Delfosse}, {Udry}, {Forveille},
  {Mayor}, {Perrier}, {Bouchy}, {Gillon}, {Lovis}, {Pepe}, {Queloz}, {Santos},
  {S{\'e}gransan}, \& {Bertaux}}]{Bonfils2013}
{Bonfils}, X., {Delfosse}, X., {Udry}, S., {et~al.} 2013, \aap, 549, A109

\bibitem[{{Buchhave} {et~al.}(2018){Buchhave}, {Bitsch}, {Johansen}, {Latham},
  {Bizzarro}, {Bieryla}, \& {Kipping}}]{Buchhave2018}
{Buchhave}, L.~A., {Bitsch}, B., {Johansen}, A., {et~al.} 2018, \apj, 856, 37

\bibitem[{{Buchhave} {et~al.}(2012){Buchhave}, {Latham}, {Johansen},
  {Bizzarro}, {Torres}, {Rowe}, {Batalha}, {Borucki}, {Brugamyer}, {Caldwell},
  {Bryson}, {Ciardi}, {Cochran}, {Endl}, {Esquerdo}, {Ford}, {Geary},
  {Gilliland}, {Hansen}, {Isaacson}, {Laird}, {Lucas}, {Marcy}, {Morse},
  {Robertson}, {Shporer}, {Stefanik}, {Still}, \& {Quinn}}]{Buchhave2012}
{Buchhave}, L.~A., {Latham}, D.~W., {Johansen}, A., {et~al.} 2012, \nat, 486,
  375

\bibitem[{{Czekaj} {et~al.}(2014){Czekaj}, {Robin}, {Figueras}, {Luri}, \&
  {Haywood}}]{Czekaj2014}
{Czekaj}, M.~A., {Robin}, A.~C., {Figueras}, F., {Luri}, X., \& {Haywood}, M.
  2014, \aap, 564, A102

\bibitem[{{Dawson} \& {Murray-Clay}(2013)}]{DawsonMurray-Clay2013}
{Dawson}, R.~I. \& {Murray-Clay}, R.~A. 2013, \apjl, 767, L24

\bibitem[{{Delfosse} {et~al.}(2013){Delfosse}, {Donati}, {Kouach},
  {H{\'e}brard}, {Doyon}, {Artigau}, {Bouchy}, {Boisse}, {Brun}, {Hennebelle},
  {Widemann}, {Bouvier}, {Bonfils}, {Morin}, {Moutou}, {Pepe}, {Udry}, {do
  Nascimento}, {Alencar}, {Castilho}, {Martioli}, {Wang}, {Figueira}, \&
  {Santos}}]{Delfosse2013}
{Delfosse}, X., {Donati}, J.-F., {Kouach}, D., {et~al.} 2013, in SF2A-2013:
  Proceedings of the Annual meeting of the French Society of Astronomy and
  Astrophysics, ed. L.~{Cambresy}, F.~{Martins}, E.~{Nuss}, \& A.~{Palacios},
  497--508

\bibitem[{{Dorn} {et~al.}(2015){Dorn}, {Khan}, {Heng}, {Connolly}, {Alibert},
  {Benz}, \& {Tackley}}]{Dorn2015}
{Dorn}, C., {Khan}, A., {Heng}, K., {et~al.} 2015, \aap, 577, A83

\bibitem[{{Drake} \& {Righter}(2002)}]{DrakeRighter2002}
{Drake}, M.~J. \& {Righter}, K. 2002, \nat, 416, 39

\bibitem[{{Fischer} \& {Valenti}(2005)}]{FischerValenti2005}
{Fischer}, D.~A. \& {Valenti}, J. 2005, \apj, 622, 1102

\bibitem[{{Gaia Collaboration} {et~al.}(2016){Gaia Collaboration}, {Prusti},
  {de Bruijne}, {Brown}, {Vallenari}, {Babusiaux}, {Bailer-Jones}, {Bastian},
  {Biermann}, {Evans}, \& et~al.}]{GaiaCollPrusti2016}
{Gaia Collaboration}, {Prusti}, T., {de Bruijne}, J.~H.~J., {et~al.} 2016,
  \aap, 595, A1

\bibitem[{{Guillot} {et~al.}(2006){Guillot}, {Santos}, {Pont}, {Iro}, {Melo},
  \& {Ribas}}]{Guillot2006}
{Guillot}, T., {Santos}, N.~C., {Pont}, F., {et~al.} 2006, \aap, 453, L21

\bibitem[{{Haywood}(2008)}]{Haywood2008}
{Haywood}, M. 2008, \aap, 482, 673

\bibitem[{{Haywood}(2009)}]{Haywood2009}
{Haywood}, M. 2009, \apjl, 698, L1

\bibitem[{{Haywood} {et~al.}(2013){Haywood}, {Di Matteo}, {Lehnert}, {Katz}, \&
  {G{\'o}mez}}]{Haywood2013}
{Haywood}, M., {Di Matteo}, P., {Lehnert}, M.~D., {Katz}, D., \& {G{\'o}mez},
  A. 2013, \aap, 560, A109

\bibitem[{{Ida} \& {Lin}(2004)}]{IdaLin2004}
{Ida}, S. \& {Lin}, D.~N.~C. 2004, \apj, 616, 567

\bibitem[{{Inaba} {et~al.}(2001){Inaba}, {Tanaka}, {Nakazawa}, {Wetherill}, \&
  {Kokubo}}]{Inaba2001}
{Inaba}, S., {Tanaka}, H., {Nakazawa}, K., {Wetherill}, G.~W., \& {Kokubo}, E.
  2001, \icarus, 149, 235

\bibitem[{{Johnson} \& {Li}(2012)}]{JohnsonLi2012}
{Johnson}, J.~L. \& {Li}, H. 2012, \apj, 751, 81

\bibitem[{{Jung} {et~al.}(2018){Jung}, {Udalski}, {Gould}, {Ryu}, {Yee}, {Han},
  {Albrow}, {Lee}, {Kim}, {Hwang}, {Chung}, {Shin}, {Zhu}, {Cha}, {Kim}, {Lee},
  {Park}, {Lee}, {Kim}, {Pogge}, {Szyma{\'n}ski}, {Mr{\'o}z}, {Poleski},
  {Skowron}, {Pietrukowicz}, {Soszy{\'n}ski}, {Koz{\l}owski}, {Ulaczyk},
  {Pawlak}, \& {Rybicki}}]{Jung2018}
{Jung}, Y.~K., {Udalski}, A., {Gould}, A., {et~al.} 2018, ArXiv e-prints
  [\eprint[arXiv]{1803.05095}]

\bibitem[{{Kuchner} \& {Seager}(2005)}]{KuchnerSeager2005}
{Kuchner}, M.~J. \& {Seager}, S. 2005, ArXiv Astrophysics e-prints
  [\eprint{astro-ph/0504214}]

\bibitem[{{Lagarde} {et~al.}(2012){Lagarde}, {Decressin}, {Charbonnel},
  {Eggenberger}, {Ekstr{\"o}m}, \& {Palacios}}]{Lagarde2012}
{Lagarde}, N., {Decressin}, T., {Charbonnel}, C., {et~al.} 2012, \aap, 543,
  A108

\bibitem[{{Lagarde} {et~al.}(2018){Lagarde}, {Reyl{\'e}}, {Robin}, {Tautvai{\v
  s}ien{\.e}}, {Drazdauskas}, {Mikolaitis}, {Minkevi{\v c}i{\= u}t{\.e}},
  {Stonkut{\.e}}, {Chorniy}, {Bagdonas}, {Miglio}, {Nasello}, {Gilmore},
  {Randich}, {Bensby}, {Bragaglia}, {Flaccomio}, {Francois}, {Korn}, {Pancino},
  {Smiljanic}, {Bayo}, {Carraro}, {Costado}, {Jim{\'e}nez-Esteban},
  {Jofr{\'e}}, {Martell}, {Masseron}, {Monaco}, {Morbidelli}, {Sbordone},
  {Sousa}, \& {Zaggia}}]{Lagarde2018}
{Lagarde}, N., {Reyl{\'e}}, C., {Robin}, A.~C., {et~al.} 2018, ArXiv e-prints
  [\eprint[arXiv]{1806.01868}]

\bibitem[{{Lagarde} {et~al.}(2017){Lagarde}, {Robin}, {Reyl{\'e}}, \&
  {Nasello}}]{Lagarde2017}
{Lagarde}, N., {Robin}, A.~C., {Reyl{\'e}}, C., \& {Nasello}, G. 2017, \aap,
  601, A27

\bibitem[{{Lodders}(2003)}]{Lodders2003}
{Lodders}, K. 2003, \apj, 591, 1220

\bibitem[{{Majewski} {et~al.}(2017){Majewski}, {Schiavon}, {Frinchaboy},
  {Allende Prieto}, {Barkhouser}, {Bizyaev}, {Blank}, {Brunner}, {Burton},
  {Carrera}, {Chojnowski}, {Cunha}, {Epstein}, {Fitzgerald}, {Garc{\'{\i}}a
  P{\'e}rez}, {Hearty}, {Henderson}, {Holtzman}, {Johnson}, {Lam}, {Lawler},
  {Maseman}, {M{\'e}sz{\'a}ros}, {Nelson}, {Nguyen}, {Nidever}, {Pinsonneault},
  {Shetrone}, {Smee}, {Smith}, {Stolberg}, {Skrutskie}, {Walker}, {Wilson},
  {Zasowski}, {Anders}, {Basu}, {Beland}, {Blanton}, {Bovy}, {Brownstein},
  {Carlberg}, {Chaplin}, {Chiappini}, {Eisenstein}, {Elsworth}, {Feuillet},
  {Fleming}, {Galbraith-Frew}, {Garc{\'{\i}}a}, {Garc{\'{\i}}a-Hern{\'a}ndez},
  {Gillespie}, {Girardi}, {Gunn}, {Hasselquist}, {Hayden}, {Hekker}, {Ivans},
  {Kinemuchi}, {Klaene}, {Mahadevan}, {Mathur}, {Mosser}, {Muna}, {Munn},
  {Nichol}, {O'Connell}, {Parejko}, {Robin}, {Rocha-Pinto}, {Schultheis},
  {Serenelli}, {Shane}, {Silva Aguirre}, {Sobeck}, {Thompson}, {Troup},
  {Weinberg}, \& {Zamora}}]{Majewski2017}
{Majewski}, S.~R., {Schiavon}, R.~P., {Frinchaboy}, P.~M., {et~al.} 2017, \aj,
  154, 94

\bibitem[{{Miglio} {et~al.}(2017){Miglio}, {Chiappini}, {Mosser}, {Davies},
  {Freeman}, {Girardi}, {Jofr{\'e}}, {Kawata}, {Rendle}, {Valentini},
  {Casagrande}, {Chaplin}, {Gilmore}, {Hawkins}, {Holl}, {Appourchaux},
  {Belkacem}, {Bossini}, {Brogaard}, {Goupil}, {Montalb{\'a}n}, {Noels},
  {Anders}, {Rodrigues}, {Piotto}, {Pollacco}, {Rauer}, {Prieto}, {Avelino},
  {Babusiaux}, {Barban}, {Barbuy}, {Basu}, {Baudin}, {Benomar}, {Bienaym{\'e}},
  {Binney}, {Bland-Hawthorn}, {Bressan}, {Cacciari}, {Campante}, {Cassisi},
  {Christensen-Dalsgaard}, {Combes}, {Creevey}, {Cunha}, {Jong}, {Laverny},
  {Degl'Innocenti}, {Deheuvels}, {Depagne}, {Ridder}, {Matteo}, {Mauro},
  {Dupret}, {Eggenberger}, {Elsworth}, {Famaey}, {Feltzing}, {Garc{\'{\i}}a},
  {Gerhard}, {Gibson}, {Gizon}, {Haywood}, {Handberg}, {Heiter}, {Hekker},
  {Huber}, {Ibata}, {Katz}, {Kawaler}, {Kjeldsen}, {Kurtz}, {Lagarde},
  {Lebreton}, {Lund}, {Majewski}, {Marigo}, {Martig}, {Mathur}, {Minchev},
  {Morel}, {Ortolani}, {Pinsonneault}, {Plez}, {Moroni}, {Pricopi},
  {Recio-Blanco}, {Reyl{\'e}}, {Robin}, {Roxburgh}, {Salaris}, {Santiago},
  {Schiavon}, {Serenelli}, {Sharma}, {Aguirre}, {Soubiran}, {Steinmetz},
  {Stello}, {Strassmeier}, {Ventura}, {Ventura}, {Walton}, \&
  {Worley}}]{Miglio2017}
{Miglio}, A., {Chiappini}, C., {Mosser}, B., {et~al.} 2017, Astronomische
  Nachrichten, 338, 644

\bibitem[{{Nayakshin} \& {Fletcher}(2015)}]{NayakshinFletcher2015}
{Nayakshin}, S. \& {Fletcher}, M. 2015, \mnras, 452, 1654

\bibitem[{{Pollack} {et~al.}(1996){Pollack}, {Hubickyj}, {Bodenheimer},
  {Lissauer}, {Podolak}, \& {Greenzweig}}]{Pollack1996}
{Pollack}, J.~B., {Hubickyj}, O., {Bodenheimer}, P., {et~al.} 1996, \icarus,
  124, 62

\bibitem[{{Pontoppidan} {et~al.}(2014){Pontoppidan}, {Salyk}, {Bergin},
  {Brittain}, {Marty}, {Mousis}, \& {{\"O}berg}}]{Pontoppidan2014}
{Pontoppidan}, K.~M., {Salyk}, C., {Bergin}, E.~A., {et~al.} 2014, Protostars
  and Planets VI, 363

\bibitem[{{Rauer} {et~al.}(2014){Rauer}, {Catala}, {Aerts}, {Appourchaux},
  {Benz}, {Brandeker}, {Christensen-Dalsgaard}, {Deleuil}, {Gizon}, {Goupil},
  {G{\"u}del}, {Janot-Pacheco}, {Mas-Hesse}, {Pagano}, {Piotto}, {Pollacco},
  {Santos}, {Smith}, {Su{\'a}rez}, {Szab{\'o}}, {Udry}, {Adibekyan}, {Alibert},
  {Almenara}, {Amaro-Seoane}, {Eiff}, {Asplund}, {Antonello}, {Barnes},
  {Baudin}, {Belkacem}, {Bergemann}, {Bihain}, {Birch}, {Bonfils}, {Boisse},
  {Bonomo}, {Borsa}, {Brand{\~a}o}, {Brocato}, {Brun}, {Burleigh}, {Burston},
  {Cabrera}, {Cassisi}, {Chaplin}, {Charpinet}, {Chiappini}, {Church},
  {Csizmadia}, {Cunha}, {Damasso}, {Davies}, {Deeg}, {D{\'{\i}}az}, {Dreizler},
  {Dreyer}, {Eggenberger}, {Ehrenreich}, {Eigm{\"u}ller}, {Erikson}, {Farmer},
  {Feltzing}, {de Oliveira Fialho}, {Figueira}, {Forveille}, {Fridlund},
  {Garc{\'{\i}}a}, {Giommi}, {Giuffrida}, {Godolt}, {Gomes da Silva},
  {Granzer}, {Grenfell}, {Grotsch-Noels}, {G{\"u}nther}, {Haswell}, {Hatzes},
  {H{\'e}brard}, {Hekker}, {Helled}, {Heng}, {Jenkins}, {Johansen},
  {Khodachenko}, {Kislyakova}, {Kley}, {Kolb}, {Krivova}, {Kupka}, {Lammer},
  {Lanza}, {Lebreton}, {Magrin}, {Marcos-Arenal}, {Marrese}, {Marques},
  {Martins}, {Mathis}, {Mathur}, {Messina}, {Miglio}, {Montalban}, {Montalto},
  {Monteiro}, {Moradi}, {Moravveji}, {Mordasini}, {Morel}, {Mortier},
  {Nascimbeni}, {Nelson}, {Nielsen}, {Noack}, {Norton}, {Ofir}, {Oshagh},
  {Ouazzani}, {P{\'a}pics}, {Parro}, {Petit}, {Plez}, {Poretti}, {Quirrenbach},
  {Ragazzoni}, {Raimondo}, {Rainer}, {Reese}, {Redmer}, {Reffert},
  {Rojas-Ayala}, {Roxburgh}, {Salmon}, {Santerne}, {Schneider}, {Schou},
  {Schuh}, {Schunker}, {Silva-Valio}, {Silvotti}, {Skillen}, {Snellen}, {Sohl},
  {Sousa}, {Sozzetti}, {Stello}, {Strassmeier}, {{\v S}vanda}, {Szab{\'o}},
  {Tkachenko}, {Valencia}, {Van Grootel}, {Vauclair}, {Ventura}, {Wagner},
  {Walton}, {Weingrill}, {Werner}, {Wheatley}, \& {Zwintz}}]{Rauer2014}
{Rauer}, H., {Catala}, C., {Aerts}, C., {et~al.} 2014, Experimental Astronomy,
  38, 249

\bibitem[{{Robin} {et~al.}(2014){Robin}, {Reyl{\'e}}, {Fliri}, {Czekaj},
  {Robert}, \& {Martins}}]{Robin2014}
{Robin}, A.~C., {Reyl{\'e}}, C., {Fliri}, J., {et~al.} 2014, \aap, 569, A13

\bibitem[{{Santos} {et~al.}(2017){Santos}, {Adibekyan}, {Dorn}, {Mordasini},
  {Noack}, {Barros}, {Delgado-Mena}, {Demangeon}, {Faria}, {Israelian}, \&
  {Sousa}}]{Santos2017}
{Santos}, N.~C., {Adibekyan}, V., {Dorn}, C., {et~al.} 2017, \aap, 608, A94

\bibitem[{{Santos} {et~al.}(2004){Santos}, {Israelian}, \&
  {Mayor}}]{Santos2004}
{Santos}, N.~C., {Israelian}, G., \& {Mayor}, M. 2004, \aap, 415, 1153

\bibitem[{{Schneider} {et~al.}(2011){Schneider}, {Dedieu}, {Le Sidaner},
  {Savalle}, \& {Zolotukhin}}]{Schneider2011}
{Schneider}, J., {Dedieu}, C., {Le Sidaner}, P., {Savalle}, R., \&
  {Zolotukhin}, I. 2011, \aap, 532, A79

\bibitem[{{Sotin} {et~al.}(2007){Sotin}, {Grasset}, \& {Mocquet}}]{Sotin2007}
{Sotin}, C., {Grasset}, O., \& {Mocquet}, A. 2007, \icarus, 191, 337

\bibitem[{{Sousa} {et~al.}(2011){Sousa}, {Santos}, {Israelian}, {Mayor}, \&
  {Udry}}]{Sousa2011}
{Sousa}, S.~G., {Santos}, N.~C., {Israelian}, G., {Mayor}, M., \& {Udry}, S.
  2011, \aap, 533, A141

\bibitem[{{Su{\'a}rez-Andr{\'e}s} {et~al.}(2017){Su{\'a}rez-Andr{\'e}s},
  {Israelian}, {Gonz{\'a}lez Hern{\'a}ndez}, {Adibekyan}, {Delgado Mena},
  {Santos}, \& {Sousa}}]{Suarez_Andres2017}
{Su{\'a}rez-Andr{\'e}s}, L., {Israelian}, G., {Gonz{\'a}lez Hern{\'a}ndez},
  J.~I., {et~al.} 2017, \aap, 599, A96

\bibitem[{{Thorngren} {et~al.}(2016){Thorngren}, {Fortney}, {Murray-Clay}, \&
  {Lopez}}]{Thorngren2016}
{Thorngren}, D.~P., {Fortney}, J.~J., {Murray-Clay}, R.~A., \& {Lopez}, E.~D.
  2016, \apj, 831, 64

\bibitem[{{Valencia} {et~al.}(2007){Valencia}, {Sasselov}, \&
  {O'Connell}}]{Valencia2007}
{Valencia}, D., {Sasselov}, D.~D., \& {O'Connell}, R.~J. 2007, \apj, 665, 1413

\bibitem[{{Wilson} {et~al.}(2018){Wilson}, {Teske}, {Majewski}, {Cunha},
  {Smith}, {Souto}, {Bender}, {Mahadevan}, {Troup}, {Allende Prieto},
  {Stassun}, {Skrutskie}, {Almeida}, {Garc{\'{\i}}a-Hern{\'a}ndez}, {Zamora},
  \& {Brinkmann}}]{Wilson2018}
{Wilson}, R.~F., {Teske}, J., {Majewski}, S.~R., {et~al.} 2018, \aj, 155, 68

\bibitem[{{Zhu} {et~al.}(2016){Zhu}, {Wang}, \& {Huang}}]{Zhu2016}
{Zhu}, W., {Wang}, J., \& {Huang}, C. 2016, \apj, 832, 196

\end{thebibliography}

\end{document}